\documentclass[preprint,aps]{revtex4}
\usepackage{epsfig}
\usepackage{graphicx}
\usepackage{wrapfig}
\def\ga{\alpha}

\def\ge{\epsilon}

\def\gd{\delta}

\def\gm{\mu}

\def\gp{\pi}
\def\gP{\Pi}

\def\gs{\sigma}

\def\gl{\lambda}
\def\gL{\Lambda}

\def\delp{\partial_+}
\def\delm{\partial_-}

\def\delmu{\partial_\gm}

\def\part{\partial}
\def\hlf{\frac{1}{2}}
\def\A0{A^{+}_0}

\def\xpl{x^{+}}
\def\xmin{x^{-}}
\def\ymin{y^{-}}

\def\ulip{\underline{p}}
\newcommand{\nc}{\newcommand}
\nc{\intl}{\int\limits_{-L}^{+L}\!dx^-}
\nc{\intly}{\int\limits_{-L}^{+L}\!{{dy^-}\over\!2}}
\nc{\zmint}{\int\limits_{-L}^{+L}\!{{dx^-}\over{\!2L}}}
\nc{\intp}{\int\limits_{0}^{+\infty}\!{{dp^+}\over\!4\gp}}
\def\beq{\begin{equation}}
\def\eeq{\end{equation}}
\def\bea{\begin{eqnarray}}
\def\eea{\end{eqnarray}}
\begin{document}
\title{Spontaneous symmetry breaking in light front field theory}
\medskip
\author{L$\!\!$'ubom\'{\i}r Martinovi\v c\\ 
Institute of Physics, Slovak Academy of Sciences \\
D\'ubravsk\'a cesta 9, 845 11 Bratislava, Slovakia } 
\thanks{Email address: fyziluma@savba.sk}

\begin{abstract}

A semiclassical picture of spontaneous symmetry breaking in light  
front field theory is formulated. It is based on a finite-volume 
quantization of self-interacting scalar fields obeying antiperiodic 
boundary conditions. This choice avoids a necessity to solve the zero 
mode constraint and enables one to define unitary operators 
which shift scalar field by a constant. The operators simultaneously 
transform the light-front Fock vacuum to coherent states with lower 
energy than the Fock vacuum and with non-zero expectation value of the scalar 
field. The new vacuum states are non-invariant under 
the discrete or continuous symmetry of the Hamiltonian. Spontaneous 
symmetry breaking is described in this way in the two-dimensional 
$\lambda\phi^4$ theory and in the three-dimensional $O(2)$-symmetric 
sigma model. A qualitative treatment of topological kink solutions    
in the first model and a derivation of the Goldstone theorem in the second 
one is given. Symmetry breaking in the case of periodic boundary conditions 
is also briefly discussed.  
\end{abstract}
\maketitle

\section{Introduction}

Spontaneous symmetry breaking is a fundamental non-perturbative 
phenomenon of quantum field theory. It occurs when the Hamiltonian of a 
theory is symmetric under a group of transformations while the ground 
state is  non-invariant. For continuous symmetries it follows that 
there exists a field operator (elementary or composite) with non-zero 
expectation value in this vacuum state.  As a consequence, the spectrum of 
such a theory contains a massless state, the Nambu-Goldstone boson 
\cite{Gold,Nambu,GSW}, if the space dimension is greater than one \cite{MW,Col}. 
This overall picture of the broken phase is well understood in the conventional 
field theory which parametrizes the space-time by means of the four-vector 
$x^\mu=(t,x,y,z)$.

On the other hand, spontaneous symmetry breaking (SSB) still remains a bit 
mysterious in the light-front (LF) 
field theory, which is defined by the choice of the LF variables $x^\gm= (\xpl,
\xmin,x,y), x^\pm=x^0 \pm x^3$ and by quantization on a surface of constant 
LF time $\xpl$. The main reason for difficulties with obtaining a  
clear picture of SSB in the LF theory is that due to positivity of the LF 
momentum operator $P^+$ the vacuum state of the interacting LF theory  
coincides with the free Fock vacuum if independent Fourier modes carrying 
$p^+=0$ (dynamical LF zero modes) can be neglected. This simplicity of the 
vacuum state is very useful in bound-state calculations but it appears to be 
problematic in other nonperturbative issues because it prohibits any vacuum 
structure in continuum LF theories where dynamical zero modes seem indeed to 
be negligible. It is often believed that the vacuum aspects enter into the LF 
theory via non-dynamical constrained zero modes which are in principle 
obtained as solutions of corresponding constraint equations. 

The present work is based on a different concept: the ``trivial" LF vacuum, 
being a simple but rigorously defined non-perturbative state, is viewed as an 
intermediate construction, not the ultimate physical vacuum state. It can 
often be systematically transformed into more complex objects by unitary 
operators that implement a symmetry of a given field theoretic model. These 
operators are well-defined (at least with a cutoff on number of field modes) 
in an infrared-regularized formulation --  quantization in a finite volume 
(or on a line of length $L$ in two dimensions) with fields (anti)periodic in 
space coordinates. Large gauge transformations and chiral symmetry are two 
examples of this approach \cite{lm01,ljprd}. A similar treatment for scalar 
field theories has not been given so far. The reason was that symmetry 
generators in a scalar theory always annihilate the LF Fock vacuum because 
due to positivity of the momentum $P^+$ they cannot contain terms composed of 
purely creation operators \cite{JS,Leut} if there are no dynamical zero modes 
in the theory. Without such terms it is not possible to transform the LF Fock 
vacuum into a more complex object and therefore one cannot construct multiple 
vacua which are a necessary condition for any SSB. A simple observation 
that underlies the present work is that for scalar theories with polynomial 
self-interaction and negative quadratic term the LF Fock vacuum is not 
the state of minimum LF energy \cite{sugitan}. The energy is lower in a  
specific coherent state and this state is {\it not} annihilated 
by the symmetry generators. Hence, the unitary operators implementing 
the discrete or continuous symmetry will generate, when applied to this 
state, a discrete or continuous set of new (semiclassical) vacuum states.   

Light front versions of SSB have been studied by a few groups of authors.  
In the unbroken phase, a zero-mode coherent state vacuum has been derived in 
$\lambda \phi^4$ theory in two dimensions and used along with the 
variational method \cite{harij}. Scalar zero mode has been assumed to be an 
independent dynamical variable. If one imposes periodic boundary conditions 
(a standard choice) this mode is however a dependent variable satisfying an 
operator constraint. Approximate methods of its solution indicated a 
development of the broken phase above the critical coupling \cite{pinssb,rege}. 
The value of the critical coupling and the critical exponent $\eta$ have also 
been determined using the Haag expansion \cite{GWer}.
In the broken phase, a variational approach with a coherent state $\vert 
\alpha \rangle$ as a trial lowest-energy state was used for small coupling 
with the zero mode manifestly neglected \cite{RThorn}. Two (approximately) 
degenerate lowest-energy levels were found and the correct value of the mass 
of the lowest excitation in the broken phase (a kink) was 
extracted by minimizing the expectation value of the Hamiltonian in the 
$\vert \ga \rangle$ states subject to the constraint on the dimensionless 
momentum $K=\frac{L}{2\gp}P^+$. The subsequent DLCQ (discrete light-cone 
quantization) studies confirmed this picture in a truly non-perturbative 
calculations and led to a detailed prediction of kink and antikink mass and 
a few additional observables \cite{kak1,kak2}.  

In four dimensions, it is usually assumed that the scalar zero mode contains a  
constant piece. As a consequence, symmetry breaking is found to  
manifest itself in a rather unusual way by a non-conservation of the current 
even in the symmetry limit while the physical vacuum is identified with the 
Fock vacuum \cite{Yam,Prem}. The concept of vacuum triviality underlies also 
an approach to dynamical symmetry breaking \cite{Maed} based on a derivation 
of gap equations from the LF constraint equations.
  
In the broken phase, considered scalar models possess two or more degenerate 
minima of the classical potential. As already indicated, one might expect that 
even in the LF theory the Fock vacuum is not the true physical vacuum 
in this case and that a unitary operator could be constructed which would 
shift the scalar field $\phi(x)$ to the true minimum of the LF energy. 
Unfortunately, for $\phi(\xpl,\xmin)$ periodic in the space coordinate 
$\xmin$, such a construction is very difficult. This is due to the complicated 
non-linear operator zero-mode constraint. On the other hand, choosing 
antiperiodic boundary conditions in $\xmin$ \cite{MB} (which is a consistent 
choice for polynomial interactions with even powers of fields) allows one 
to define shift operators which transform the Fock vacuum to new states   
that correspond to lower LF energy. They are coherent states of large but finite 
number of Fourier modes. For simplicity, we will illustrate this mechanism in two 
well known low-dimensional scalar models. The first one is the two-dimensional  
$\lambda \phi^4$ theory in broken phase, possessing classically two degenerate 
ground states. The second model is a three-dimensional O(2)-symmetric linear 
sigma model. It has a continuum of degenerate vacuum states and one can expect 
the Goldstone phenomenon to take place. Both models are superrenormalizable. 
Renormalization can be performed by normal ordering the Hamiltonian or  
equivalently by adding a mass counterterm (a tadpole) in the first case and a 
tadpole together with the second-order self-energy counterterm in the second 
case \cite{Chang,Mag}.   

A short description of SSB in the case of periodic boundary conditions will 
also be given. We will show that some features of the broken phase are similar 
in the both cases. Since our approach is based on the quantization in a finite 
spatial volume with antiperiodic boundary conditions, we use the 
correspondingly defined sign function and Dirac's delta function. Their 
regularized form is displayed in the Appendix.   

\section{Spontaneous symmetry breaking in $\lambda\phi^4(1+1)$ theory}   

Let us consider two-dimensional $\lambda\phi^4$ theory in the broken phase.  
It is defined by the covariant Lagrangian density 
\beq
{\cal L}=\hlf \delmu \phi \part^ \gm \phi + \hlf \gm^2 \phi^2 - \frac{\gl}{4}
\phi^4,\;\gm^2>0,
\label{lagr}
\eeq
which is invariant under the discrete transformation of the real scalar field 
$\phi(x) \rightarrow -\phi(x)$. Classically, the potential energy in 
(\ref{lagr}) has two minima at $\phi_c=\pm \mu/\sqrt{\lambda}$. In the 
tree-level analysis, one usually shifts the field by $\pm \phi_c$ and obtains 
two Lagrangians which reveal the particle spectrum of the theory in terms of 
``small'' oscillations above $\phi_c$. The original symmetry becomes hidden 
in the sense that the two Lagrangians are individually not symmetric under 
$\phi(x) \rightarrow -\phi(x)$ but the symmetry operation transforms one to  
the other. Recall that due to the existence of more than one minimum of the 
potential, the model exhibits in addition to symmetry breaking also nontrivial 
topological properties \cite{Raj}. There exist solutions of the classical 
equations of motion with finite energy which interpolate between the minima. 
They carry a conserved topological charge, proportional to the difference 
of the field values at the boundaries, and corresponding to the conserved 
topological current $k^\rho=\frac{\sqrt{\lambda}}{\mu}\ge^{\rho \nu}\part_\nu \phi.$
 
The Lagrangian (\ref{lagr}) is expressed in terms of the LF variables as    
\bea 
{\cal L}_{lf} = 2\delp\phi\delm\phi + \hlf \gm^2\phi^2 - \frac{\gl}{4}\phi^4,
\label{lflagr}
\eea
where $\partial_{\pm}=\partial/\partial x^{\pm}$. We restrict the spatial 
coordinate by $-L \leq \xmin \leq L$. In order to obtain a clear physical 
picture of SSB it is desirable to avoid the complicated non-linear operator   
zero-mode constraint present in the case of periodic boundary conditions. 
We impose therefore the antiperiodic boundary condition 
(BC) $\phi(L)=-\phi(-L)$ which results in discrete Fourier modes 
\beq
p^+_n=\frac{2\gp}{L}n,\;\;n=1/2,3/2,\dots\infty. 
\label{newp}
\eeq
The antiperiodic BC also implies that in the quantum theory we can 
define the operator of the topological charge $Q=\frac{\sqrt{\lambda}}{\mu}[
\phi(L)-\phi(-L)]=2\frac{\sqrt{\lambda}}{\mu}\phi(L)$. 

The standard canonical treatment yields the energy-momentum tensor components 
$T^{+-}$ and $T^{++}$ which define the LF Hamiltonian $P^-$
\beq
P^- = \hlf\intl T^{+-}(\xmin) = \hlf\intl :\big[-\gm^2\phi^2 + 
\frac{\gl}{2}\phi^4 \big]:, 
\label{Ham}
\eeq
as well as the LF momentum operator 
\beq
P^+ = \hlf \intl T^{++}(\xmin) = \hlf \!\int\limits_{-L}^{L}d\xmin 
4:\big[\delm \phi \delm \phi\big]:.
\label{momo}
\eeq
The field expansion in terms of the Fourier modes at $\xpl=0$ reads
\beq
\phi(0,\xmin)=\frac{1}{\sqrt{2L}}\sum_{n=1/2}^\infty \frac{1}{\sqrt{p^+_n}}
\big[a_n e^{-\frac{i}{2}p^+_n\xmin} + a_n^\dagger e^{\frac{i}{2}p^+_n\xmin}
\big].
\label{Fexp}
\eeq 
The annihilation and creation operators are 
required to satisfy the quantization condition $[a_m,a_n^\dagger] = \gd_{mn}$. As 
a consequence, one recovers the usual commutator at equal LF times,  
\beq
\left[\phi(0,\xmin),\phi(0,\ymin)\right]=-\frac{i}{8} \ge_a(\xmin-\ymin),
\label{CR}
\eeq
where $\ge_a(\xmin)$ is the antiperiodic sign function
\beq
\ge_a(\xmin)=\frac{4i}{L}\sum_{n=1/2}^{\infty} \frac{1}{p^+_n}
\big[e^{-\frac{i}{2}p^+_n\xmin} - e^{\frac{i}{2}p^+_n\xmin} \big],
\label{signf}
\eeq
defined in terms of the discrete momenta (\ref{newp}). 
The conjugate momentum $\gP_\phi$ is not equal to the time derivative of 
the scalar field in the LF theory. It is a dependent variable, determined by  
$\phi(x)$ itself, $\gP_\phi = 2\delm \phi$ \cite{LKS}. Hence, the alternative 
form of the basic commutation relation, following from Eq.(\ref{CR}), is 
\beq
\big[\phi(0,\xmin),\gP_\phi(0,\ymin)\big] = \frac{i}{2}\gd_a(\xmin-\ymin),
\label{CR1}
\eeq
where $\gd_a(\xmin)$ is the antiperiodic delta function, $\gd_a(\xmin)=
1/2\delm\ge_a(\xmin)$. The same quantization rules can be obtained more 
rigorously  by the Dirac-Bergmann method \cite{DB} for constrained systems.  

Consider now a unitary operator 
\beq
U(b)=\exp \big[-2ib\intl \gP_\phi(\xmin) \big].
\label{trop}
\eeq
For antiperiodic boundary conditions, it reduces to  
\beq
U(b)=e^{-8ib \phi(L)}
\label{U}
\eeq
and translates the field $\phi(\xmin)$ by a constant $b$ as can be easily 
shown by using the operaor identity $\exp(A) B \exp(-A) = 
B+[A,B] + \dots$~: 
\bea
U(b)\phi(\xmin)U^{-1}(b) &=& \phi(\xmin) - 8ib\big[\phi(L),\phi(\xmin)
\big] 
\nonumber \\ 
&=& \phi(\xmin) -  b\ge_a(L - \xmin). 
\label{shift}
\eea
Thus, the antiperiodic scalar field can be shifted by a constant without 
violating its antiperiodicity. The reason for that is the simple property 
of the sign function $\ge_a(L - \xmin)$: it is equal to 1 
for all $\xmin$ in the box except for the endpoints where it drops to zero. 
This is of course a direct consequence of the basic property $\ge_a(0)=
\ge_a(2L)=0$. It is much more difficult to perform a similar shift of the 
field in the case of periodic boundary condition because of the presence 
of the a priori unknown operator zero mode. 
Recall for comparison that since in 
the conventional space-like quantization the conjugate momentum is a dynamical 
quantity, the volume integration in the shift operator analogous to 
Eq.(\ref{trop}) projects out only its zero-mode component \cite{IZ}.  

We should note however that the above considerations were a bit formal 
and the actual situation is slightly more complicated. The point is that the 
operator $U(b)$ (\ref{U}) exists (is non-zero) only if we impose a cutoff on 
the number of modes (see Eq.(\ref{CS}) and the discussion after 
Eq.(\ref{vac2})). Consequently, the sign function in (\ref{shift}) is 
replaced by a truncated series $\ge_{\gL} (\xmin)$ defined by Eq.(\ref{signf}) 
with $n \leq \gL$. 

We may use $U(b)$ to generate a family of shifted vacuum 
states $\vert b \rangle =U(b)\vert 0 \rangle$, where $\vert 0 \rangle$ is the 
Fock vacuum, $a_n\vert 0 \rangle = 0.$ Can one of these states be a better 
candidate for the true physical vacuum? To determine this, let us minimize 
the  expectation value of the LF Hamiltonian,
\beq 
\langle b \vert P^-\vert b \rangle = 
\langle 0 \vert U^{-1}(b)P^- U(b)\vert 0 \rangle= 
\langle 0 \vert \hlf \intl T^{+-}_b(\xmin) \vert 0 \rangle,  
\eeq
\label{EVEV}
where
\beq
T_b^{+-}(\xmin)= :\!\big[-\gm^2\big(\phi+b
\ge_{\gL}(L-\xmin)\big)^2 + \frac{\gl}{2} \big(\phi+b\ge_{\gL}(L-\xmin)\big)^4 \big]:
\label{Tb}
\eeq
As shown in the Appendix, for sufficiently large value of $\gL$ the function   
$\ge_{\gL}(L-\xmin)$ differs only negligibly from unity on the  interval 
$-L \le \xmin \le L$. The same is true also for its powers. We will therefore 
suppress henceforth symbol of the sign function in the formulae similar 
to \ref{Tb}.  Also, due to the finite number of Fourier modes, the function 
$\ge_\gL(L-\xmin)$ does not have an exactly rectangular shape but it is 
smooth in the neighborhood of the points $\xmin=\pm L$ (see the Appendix). 

Now, for the expectation value of the energy we find $\langle 
b \vert P^-\vert b \rangle = Lb^2(\frac{\gl}{2}b^2-\gm^2)$ 
which has a non-trivial minimum for $b^2=\frac{\gm^2}{\gl}\equiv v^2$. The LF  
energy density is lower in the new vacuum $\vert v \rangle$:  
\beq
\langle v \vert P^- \vert v \rangle/2L = -\frac{\gm^4}{4\gl} \;\;<  
\;\; \langle 0 \vert P^- \vert 0 \rangle/2L = 0. 
\label{emin}
\eeq
The vacuum expectation value (VEV) of the scalar field in this state coincides  
with the position of the minimum of the classical potential: 
\beq
\langle v \vert \phi(x)\vert v \rangle = \langle 0 \vert U^{-1}(v)
\phi(x)U(v)\vert 0 \rangle = \frac{\gm}{\sqrt{\lambda}}\ge_{\gL}(\xmin-L) 
= \frac{\gm}{\sqrt{\lambda}}.
\label{FVEV}
\eeq 
The last equality holds in the sense discussed after Eq.(\ref{Tb}). 
The fact that the field expectation value is not a perfect constant is irrelevant 
here. The crucial point is that the scalar field can be shifted by a non-operator 
piece (which is a c-number multiplied by a function approaching unity for infinite 
number of field modes).       

Inserting the field expansion (\ref{Fexp}) into the definition of $U(v)$, 
we get a coherent state representing the physical vacuum of the model in the 
semi-quantum approximation: 
\beq
\vert v \rangle =\exp\big\{v\!\! \sum_{n=1/2}^{\gL}\tilde{c}_n\big(
a_n^\dagger -  a_n \big)\big\}\vert 0 \rangle = 
{\cal N}\exp\big\{v\!\!\sum_{n=1/2}^{\gL}\tilde{c}_n a_n^\dagger\big\} 
\vert 0 \rangle, 
\label{CS}
\eeq
where
\beq
\tilde{c}_n=4(-1)^{n-1/2}/\sqrt{\gp n},
~~~{\cal N}=\exp\big\{-\frac{v^2}{2}\sum_{n=1/2}^{\gL} \tilde{c}_n^2
\big\} \approx \exp\big\{-\frac{8v^2}{\gp} \ln \gL\big\}. 
\label{CSa}
\eeq
Notice that the coherent states (\ref{CS}) are $L$-independent and also 
correctly normalized, $\langle v \vert v \rangle = 1$. Further, the scalar 
product $\langle -v \vert v \rangle = {\cal N}^4 = \gL^{-32v^2/\gp}$
and thus the overlap between the two vacua vanishes 
in the limit $\gL \rightarrow \infty$. This means that, in contrast to the 
space-like theory, the two vacua are orthogonal even in the finite volume as 
long as the number of degrees of freedom is infinite. The corresponding 
multiparticle spaces can be generated by applying creation operators 
$a^\dagger_n$ on $\vert v \rangle$. These states however do not form an 
orthogonal basis. Alternatively, one can transform the original Fock states, 
built on $\vert 0 \rangle$, by means of $U(v)$ \cite{qostates}. The Hamiltonian matrix elements will be  (up to normalization) of the form 
\beq
\langle 0 \vert a_{m_1}a_{m_2}...a_{m_i}U^{-1}(v)P^-U(v)a^\dagger_{n_j}
 ...a^\dagger_{n_2} a^\dagger_{n_1} \vert 0 \rangle. 
\label{vham}
\eeq
%
In both cases the physically relevant Hamiltonian is the transformed 
(``effective'') one, equal to $P^-_{(v)} = U^{-1}(v)P^-U(v)$: 
\beq
P^-_{(v)} = \hlf \intl:\big[2\gm^2\phi^2 +\frac{\gl}{2}\phi^4 + 
2\gl v \phi^3  -  \frac{\gm^4}{2\gl}\big]:.
\label{EP}
\eeq
It has a correct sign of the term quadratic in $\phi$ and thus describes a 
massive scalar field with mass equal to $\sqrt{2}\gm$. However, it has lost  
the symmetry of the original Hamiltonian under $\phi(x) \rightarrow 
-\phi(x)$ --   this symmetry has been broken by choosing $\vert v \rangle$ 
as the vacuum state. Actually, the theory originally had also the second 
ground state. This one can demonstrate by considering      
a unitary operator that implements the original discrete symmetry,   
\beq
V(\gp)=\exp \big[-i\gp\!\sum_{n=1/2}^\gL a_n^\dagger a_n \big].
\label{V}
\eeq
It acts correctly on the creation and annihilation operators, 
\beq
V(\gp) a_n V^-(\gp)=-a_n,\; V(\gp) a^\dagger_n V^-(\gp)=
-a^\dagger_n
\label{parity}
\eeq
and hence leaves $P^-$ invariant, $V(\gp)P^-V^-(\gp) = P^-$. 
The operator $V(\gp)$ generates the second vacuum:
\beq
V(\gp)\vert v \rangle = \vert - v \rangle,
\label{vac2}
\eeq
since $V(\gp)U(v)=U(-v)V(\gp)$. The latter relation follows 
from the operator identity $\exp(A) \exp(B) = 
\exp(e^\rho B)\exp(A)$, 
valid if $[A,B]=\rho B$ ($\rho=$ real parameter.) We easily find   
$\langle -v \vert \phi(\xmin)\vert -v \rangle = -v$. The corresponding 
``effective'' Hamiltonian $P^-_{(-v)}$ in the space sector built on 
$\vert -v \rangle$ coincides with the expression (\ref{EP}) up to the opposite 
sign of the cubic term. Although both Hamiltonians are individually not 
invariant, they are connected by the "parity" transformation: $P^-_{(-v)} = 
V(\gp)P^-_{(v)} V^{-1}(\gp)$ and vice versa. We can choose 
any of the two vacua and their corresponding ``effective'' Hamiltonian to 
describe the physical system under study.    

An alternative way of obtaining the coherent state vacuum (\ref{CS}) is to 
minimize the expectation value of the Hamiltonian in the coherent 
states $\vert \alpha \rangle$, $\vert \ga\rangle \sim \exp{\big(
\sum\ga_na^\dagger_n\big)}\vert 0 \rangle$, imposing the condition that 
the expectation value of the antiperiodic field is constant. If one  
requires instead of a constant value for $\langle \alpha\vert \phi(\xmin)
\vert \alpha\rangle$ the value $-v$ for $-L\le \xmin \le 0$ and $v$ for 
$0 \le \xmin \le L$, i.e. a step-like shape, one obtains a configuration 
that also minimizes the LF energy and qualitatively approximates a kink 
\cite{kak1}:
\bea
\vert \ga \rangle = \exp\big[v \sum_{n=1/2}^{\gL}
\ga_n\big(a_n^\dagger - a_n \big)\big]\vert 0 \rangle,~~~ 
\ga_n =\frac{4i}{\sqrt{\gp n}}.
\label{kinkst}
\eea
In $x$-representation, the state $\vert \ga \rangle$ can be expressed in terms 
of the unitary operator $W(v)$ as
\beq
\vert \ga \rangle = W(v)\vert 0 \rangle ,~~W(v)=e^{i8v\phi(0)}
\label{W}
\eeq
and one easily obtains
\beq
\langle \ga\vert \phi(x)\vert \ga \rangle = \langle 0\vert W^{-1}(v)
\phi(x)W(v)\vert 0 \rangle = v\ge_\gL(\xmin),
\label{kink}
\eeq
which is the result indicated above.
Note also that the kink state $\vert \ga \rangle$ is for $\gL \rightarrow 
\infty$ orthogonal to the vacuum 
state \cite{GJ}, $\langle v \vert \ga \rangle \sim \exp \big(-\ln\gL \big)$. 
These states belong 
to the sectors with different topological charges (superselection sectors):
\bea
&&\langle \ga \vert Q \vert \ga \rangle = v^{-1}\langle 0 \vert W^{-1}(v)\phi(L)
W(v)\vert 0 \rangle = 8i[\phi(L),\phi(0)]=\ge_\gL(L)= 1.  
\nonumber \\
&&\langle v \vert Q\vert v \rangle = v^{-1}\langle 0 \vert U^{-1}(v)\phi(L)U(v)
\vert 0 \rangle = v^{-1}\langle 0 \vert \phi(L) \vert 0 \rangle = 0. 
\label{topova}
\eea
Quantitative predictions of the properties of kink and antikink 
in quantum theory were obtained by LF Hamiltonian matrix diagonalizations 
using discretized light cone quantization \cite{RThorn,kak1}.   

Finally, let us discuss the LF momentum of the coherent-state vacuum $U(v)\vert 
0 \rangle$ and of the transformed Fock states $U(v)a^\dagger_{m_1}a^\dagger_
{m_2}...\vert 0 \rangle$. Our vacua are not momentum eigenstates since they are by definition only eigenstates of the annihilation operator. They represent an 
approximation to the true physical vacuum. One can calculate expectation 
values of physical quantities in these states. The VEV of an unordered $P^+$ 
would be 
\beq 
\langle v \vert P^+ \vert v \rangle = 2\intl \langle 0 \vert\big[\delm\big(
\phi(x) + v\ge_\gL(L-\xmin)\big)\big]^2\vert 0 \rangle = 
\frac{\gp}{L}\sum_{n=1/2}^{\gL}
\big(n + \frac{32}{\gp}v^2 \big).
\label{MVEV}
\eeq
The first term on the right-hand side is removed by normal ordering. 
The second term, equal to $16 v^2 \gd_\gL(0)$ is a consequence of the 
fact that $\delm \ge_\gL(L - \xmin) = - 2 \gd_\gL(L - \xmin)$ which for 
$\gL \rightarrow \infty$ is singular just at the endpoints $\xmin=\pm L$. 
For finite $\gL$ the second term is a finite constant $C$. It is also 
present in the expectation values of the LF momentum of particle 
states:
\bea
&&\langle 0 \vert a_{l}U^{-1}(v)P^+ U(v) a^\dagger_l\vert 0 \rangle = 
p^+_l + C, 
\nonumber \\
&&\langle 0 \vert a_{k}a_{l}U^{-1}(v)P^+ U(v) a^\dagger_{k} a^\dagger_l\vert 0 
\rangle = p^+_k + p^+_l + C, 
\label{onep}
\eea
and similarly for higher many-particle states. Thus the LF momentum of the 
transformed states is shifted by the same constant value which is 
physically irrelevant since it cancels in the differences between 
any two levels. We shall therefore subtract this unphysical 
constant. 
Let us remark that the necessity to perform the (trivial) renormalization of  
the $P^+$ operator may seem a little unusual but actually it is natural and 
physically transparent: the shift of the scalar field due to $U(v)$ is almost 
precisely equal to a constant in the whole box except for the small 
neighbourhood of the endpoints. 
The expectation values of the momentum operator receive large but common 
contributions from the neighbourhood of the endpoints due to an $\xmin$- 
integral over $[\gd_\gL(L-\xmin)]^2$.   

Since the approximative vacuum states $\vert v\rangle$ are not eigenstates of 
$P^+$, the translational invariance of the theory can only be formulated in a 
weaker form. The Heisenberg equation $-2i\delm \phi(x)=[P^+,\phi(x)]$ is 
satisfied on the vacuum state in the sense of matrix elements. The usual 
condition $\exp{\big(i\vec{a}.\vec{P}\big)} \vert vac \rangle = \vert vac 
\rangle$ implying $\langle vac \vert \exp{\big(i\vec{a}.\vec{P}\big)} \vert 
vac \rangle = 1$ is replaced by $\langle v \vert \exp{\big(\frac{i}{2}a^-P^+
\big)} \vert v \rangle = \exp{\big(\frac{i}{2}a^-C\big)}$ here. 
 
Our formulation of SSB in the two-dimensional scalar model can be used 
as a basis for studying phase transition to unbroken phase by means of  
Hamiltonian matrix diagonalization \cite{Dean,DHJ,James} (the DLCQ method). 
We expect that new matrix elements (\ref{vham}) generated by working with 
the vacuum $\vert v \rangle$ and the Hamiltonian $P^-_v$ 
will be important for a correct description of the phase transition which 
should occur if one varies the coupling constant keeping the mass parameter 
fixed \cite{Chang,Mag}. An improvement of the computations of the LF 
energy eigenstates \cite{kak1} with the DLCQ method which led to a direct 
evidence of topological excitations, can be envisaged, too. One may hope to  
obtain also quantum corrections to the semiclassical coherent-state vacuum 
(\ref{CS}). Recall that in the DLCQ method one diagonalizes Hamiltonian 
matrices for a fixed value of the dimensionless momentum 
$K=\frac{L} {2\gp}P^+$. The value of $K$ determines simultaneously the maximum 
momentum mode in the Fock expansion of the field and hence the summation in the 
coherent-state vacuum (\ref{CS}) will be truncated by $K$. The corresponding 
transformed DLCQ Hamiltonian $\tilde{H}=\frac{2\gp}{L}P^-$ takes the form
\bea 
&\tilde{H}=-\mu^2\sum\limits_{n=\hlf}^\gl \frac{1}{n}A^\dagger_n A_n + 
\frac{\gl}{8\gp} \sum\limits_{klmn}^\gL \frac{1}{\sqrt{klmn}} 
\Big[ 2 A^\dagger_k A_l A_m A_n \gd_{k,l+m+n} + \nonumber \\
&+ 3 A^\dagger_k A^\dagger_l A_m A_n \gd_{k+l,m+n} 
+ 2 A^\dagger_k A^\dagger_l A^\dagger_m A_n \gd_{k+l+m,n} \Big], \nonumber \\
&A_n=U^{-1}(v)a_nU(v)=a_n + v\tilde{c}_n.
\label{DLCQH}
\eea
For large $\gL$ this Hamiltonian approaches the limiting form (\ref{EP}) as 
can be shown by evaluating explicitly the powers of the operators $A_n$, 
regrouping terms and using the definition of the sign function in terms of 
discrete modes. In real DLCQ computations one should diagonalize 
the above Hamiltonian calculated in the Fock basis for given value of 
$K \approx 40-60$ and then extrapolate results to $K \rightarrow \infty$. 

\subsection{SSB with periodic boundary conditions}

Previous attempts to understand SSB in the LF theory were made either 
without imposing boundary conditions explicitly or by employing periodic ones 
\cite{Tsugi}, typically starting from the symmetric phase of the theory. Can 
one give a formulation of the broken phase using PBC? The problem is 
complicated because one has to solve the operator constraint for the dependent 
zero mode $\phi_0$. At present, this appears possible only for small coupling, 
where one can use perturbation theory. Perturbative solution is however quite 
interesting because it corresponds to the semiclassical regime of the broken 
phase and one can compare the results with the results of the previous 
section. The physical picture obtained by imposing antiperiodic boundary 
condition should be quite accurate far from  the critical region, i.e. for 
small value of the coupling constant. Since a derivation of a semiclassical 
vacuum state similar to the case of antiperiodic boundary conditions seems 
not to be possible for PBC, one may expect that the physical vacuum state will 
coincide with the Fock vacuum and SSB will manifest itself by the presence of 
two Hamiltonians \cite{pinssb}. 

The field equation for the scalar field following from the Lagrangean 
(\ref{lflagr}) is 
\beq
4 \delp\delm\phi = \mu^2\phi + \gl\phi^3. 
\label{feq}
\eeq
The scalar field can be decomposed as $\phi(x)=\phi_0(\xpl) +  
\varphi(\xpl,\xmin)$, with $\phi_0$ being the $\xmin$- independent part  
carrying $p^+=0$. Projection of the field equation (\ref{feq}) on the 
zero-mode sector 
\beq
\mu^2\phi_0=-\gl \intl \big(\phi_0+\varphi\big)^3
\label{zmc}
\eeq
shows that $\phi_0$ is a dependent variable which has to be expressed in 
terms of all other (normal) modes \cite{MY}. The perturbative solution of 
the classical zero mode constraint was given by Robertson \cite{Dave} and 
has two physical branches: 
\bea
\phi_0^{(1)} &=& \frac{\mu}{\sqrt{\gl}} -\frac{3}{2}{\frac{\sqrt{\gl}}{\mu}}
\zmint \varphi^2 -\hlf\frac{\gl}{\mu^2}\zmint \varphi^3 \nonumber \\ 
\phi_0^{(2)} &=& -\frac{\mu}{\sqrt{\gl}} + \frac{3}{2}{\frac{\sqrt{\gl}}{\mu}}
\zmint \varphi^2 -\hlf\frac{\gl}{\mu^2}\zmint \varphi^3.
\label{zmsol}
\eea
To the given order it can be taken over to the quantum theory since there is 
no ordering ambiguity. Note that the solutions contain a constant piece and 
their structure differs completely from the perturbative solution in the 
symmetric phase \cite{yamdur} because of the opposite sign of the $\mu^2$-term 
in the field equation. Under $\varphi \rightarrow -\varphi$, we have 
$\phi_0^{(1)} 
\rightarrow -\phi_0^{(2)}$ and vice versa. When these two solutions are 
inserted into the PBC Hamiltonian, analogous to (\ref{Ham}), 
\beq
P^-=\hlf \intl\big[-\mu^2 \big(\phi_0 + \varphi \big)^2 + \frac{\gl}{2} 
\big(\phi_0+ \varphi \big)^2 \big],
\label{PBCHam}
\eeq
one indeed gets through $O(\gl)$ two Hamiltonians
\beq
P^-_{(\pm v)} = \hlf \intl \Big[2\mu^2\varphi^2 + \frac{\gl}{2}\varphi^4 
\pm 2\mu\sqrt{\gl}
\varphi^3 - \frac{\mu^4}{2\gl} - \frac{9}{2}\gl\varphi^2\zmint\varphi^2 \Big].
\label{2Ham}
\eeq
Their structure is similar to the Hamiltonians $P^-_v$ from the case 
of antiperiodic boundary conditions. Each Hamiltonian separately violates 
the symmetry under $\varphi \rightarrow -\varphi$ but the transformation 
connects them. Any of them can be chosen for calculating physical properties 
of the system. Their eigenstates will also be connected by the "parity" 
transformation. It is an interesting problem for DLCQ to find the lowest 
energy levels of the Hamiltonians (\ref{2Ham}).

\section{$\!\!\!\!\!\!\!$
Symmetry breaking in $\lambda(\phi^*\phi)^2(2+1)$ theory}

As the next step, we could consider a two-dimensional theory of a  
self-interacting complex scalar field. The corresponding Hamiltonian has a 
continuous symmetry instead of the discrete one. The full treatment requires 
a discussion of the LF version of the Coleman theorem which prohibits SSB in 
one space dimension \cite{Col}. Since this topic deserves a separate analysis,  
here we will study the $O(2)$ symmetric sigma model in three  
dimensions. It is defined by the classical Lagrangean 
density
\beq
{\cal L} = \hlf\delmu\phi^\dagger \partial^\gm \phi + \hlf \gm^2 \phi^\dagger
\phi - \frac{1}{4}\gl(\phi^\dagger \phi)^2.
\label{cphil}
\eeq
The system will be studied in a finite volume $V=4LL_\perp$,  
$-L \leq \xmin \leq L,~ -L_\perp \leq x_\perp \leq L_\perp$. Scalar    
fields are taken antiperiodic in both $\xmin$ and the transverse  
coordinate $x_\perp$. In terms of two real scalar fields introduced by 
$\phi(x)=\gs(x)+i\gp(x)$, the corresponding LF Lagrangean density  
\bea
{\cal L}_{lf} &=&  2\delp\gs\delm \gs + 2 \delp \gp \delm \gp - 
\hlf(\part_\perp \gs)^2 - \hlf (\part_\perp \gp)^2 + \nonumber \\ 
&+&\frac{\gm^2}{2}\big(\gs^2 + \gp^2 \big)  -  
\frac{\gl}{4}(\gs^2+\gp^2)^2 
\label{Lsipi}
\eea 
is invariant under $O(2)$ rotations
\bea
\gs(x) \rightarrow \gs(x) \cos \ga  - \gp(x) \sin \ga , \nonumber \\
\gp(x) \rightarrow  \gs(x) \sin \ga  +  \gp(x) \cos \ga .
\label{rotsipi}
\eea
The associated conserved current is $j^\gm = \gs \part^\gm 
\gp - \part^\gm \gs \gp$. The field expansions at $\xpl=0$ are 
\beq
\gs(\underline{x})=\frac{1}{\sqrt{V}}\sum_{\underline{n}} \frac{1}
{\sqrt{p^+_n}} \big[a(p_{\underline{n}}) e^{-i p_{\underline{n}}.\underline{x}}
 + a^\dagger(p_{\underline{n}}) e^{i p_{\underline{n}}.\underline{x}} \big],
\eeq

\beq
\gp(\underline{x})=\frac{1}{\sqrt{V}}\sum_{\underline{n}} \frac{1}
{\sqrt{p^+_n}} \big[c(p_{\underline{n}}) e^{-i p_{\underline{n}}.\underline{x}} 
+ c^\dagger(p_{\underline{n}}) e^{i p_{\underline{n}}.\underline{x}} \big].  
\label{spexp}
\eeq
We use the notation $\underline{x}=(\xmin,x_\perp)$, 
$\underline{n}\equiv(n,n_\perp), p_{\underline{n}}= (p^+_n,p_{n_\perp})=
(\frac{2\gp}{L}n,\frac{\gp}{L_\perp} n_\perp)$ with $n,n_\perp =1/2,3/2,\dots 
\infty$. 
The conjugate momenta are $\gP_\gs=2\delm\gs, \gP_\gp=2\delm\gp$.  
The $\gs$ field operators satisfy the commutation relation  
\bea 
\left[\gs(0,\underline{x}),\gs(0,\underline{y})\right]=
-\frac{i}{8}\ge_a(\xmin \!\!-\ymin) 
\gd_a(x_\perp\!-y_\perp).
\eea 
The commutator of the $\gp$ fields has the same form. The Hamiltonian is 
\bea  
&&P^- = \int\limits_{V}^{}\!\! d^2\underline{x} \big[
(\part_\perp\gs)^2 + (\part_\perp\gp)^2 + 2V(\gs^2+\gp^2)\big], \nonumber \\
&&V(\gs^2+\gp^2)=-\frac{\gm^2}{2}\big(\gs^2 + \gp^2 \big)+\frac{\gl}{4}
(\gs^2+\gp^2)^2,  
\label{WH}
\eea
where $d^2\underline{x}=\frac{1}{2}dx^-dx_\perp$. In principle, both $\gs(x)$ 
and $\gp(x)$ can be transformed by the unitary operators $U_\gs(b)$ and 
$U_\gp(b)$ in analogy with Eq.(\ref{shift}). It is simpler however to start by 
shifting only one field which we choose in accord with the standard 
treatment to be $\gs$: 
\beq
U_\gs(b)\gs(\underline{x})
U_\gs^{\dagger}(b)=\gs(\underline{x}) - b\ge_\gL(L-\xmin)\ge_\gL(x_\perp - 
L_\perp),
\label{sigshift}
\eeq 
with  
\bea 
U_\gs(b)=\exp \big[\!-\!4ib\!\int\limits_{V}^{}\!\!d^2\underline{x}
\gP_\gs(\underline{x}) \big] 
= \exp \big[\!-\!8ib\!\!\!\!\int\limits_{-L_\perp}^
{+L_\perp}\!\!\!\!dx_\perp \gs(L,x_\perp)\big].
\label{sishift}
\eea

By minimization of $\langle b;0 \vert P^- \vert b;0 \rangle$, where $\vert b;0  
\rangle = U_\gs(b)\vert 0 \rangle$, we find that the (approximate)physical 
vacuum $\vert v;0 \rangle = U_\gs(v) \vert 0 \rangle$ corresponds to the 
value $b=\frac {\gm}{\sqrt{\gl}} \equiv v$ and 
\bea
&&\vert v;0 \rangle = \exp\Big\{-v \sum_{\underline{n}}\tilde{c}
(p_{\underline{n}})
\big[a^\dagger(p_{\underline{n}}) - a(p_{\underline{n}})\big]\Big\}\vert 0 
\rangle,\; \label{v}\\
&&\tilde{c}(p_{\underline{n}})=\frac{8}{\gp}
\sqrt{\frac{L_\perp}{2\gp}}\frac{(-1)^{n+n_\perp}}{\sqrt{n}n_\perp}.
\label{CS3}
\eea
The rotations (\ref{rotsipi}) are implemented by the unitary operators 
$V(\ga)= e^{i\ga Q}$, where $Q=\int\limits_{V}^{} d^2\underline{x}j^+
(\underline{x})$: 
\beq
\gs(x) \rightarrow V(\ga)\gs(x)V^{\dagger}(\ga), \gp(x) \rightarrow V(\ga)
\gp(x) V^{\dagger}(\ga), 
\label{shift2}
\eeq
\beq 
V(\ga)=\exp \big[\ga\sum_{\underline{n}}\left(a^\dagger(p_
{\underline{n}}) c(p_{\underline{n}}) - c^\dagger(p_{\underline{n}})
a(p_{\underline{n}})\right)\big].
\label{Frot}
\eeq
The operators $V(\ga)$ extend the ``primary'' vacuum $\vert v;0 \rangle$ 
to the infinite family $\vert v;\ga\rangle = V(\ga)\vert v \rangle$. 
Explicitly, we get  
\beq
\vert \ga;v \rangle =  
\exp\Big\{\!-\!v \sum_{\underline{n}}\tilde{c} (p_{\underline{n}})
\big[\big(a^\dagger(p_{\underline{n}}) - a(p_{\underline{n}})\big)\cos \ga 
\!+\! \big(c^\dagger(p_{\underline{n}})- c(p_{\underline{n}})\big)\sin \ga \big]
\Big\} \vert 0 \rangle.
\label{genvac}
\eeq 
In spite of the presence of the box length $L_\perp$ in the coherent state   
(\ref{v}), the orthogonality $\langle v; \ga \vert v;\ga^\prime \rangle 
= \gd_{\ga\ga^\prime}$ holds in the limit of infinite number of 
longitudinal modes $n$.
 
We can interpret the relation for the vacuum and particle matrix elements of 
$P^-$ (cf. Eq.(\ref{vham})) as defining an effective Hamiltonian  
$P^-_v = U^{\dagger}_\gs(v)P^- U_\gs(v) $: 
\bea
&P^-_v = \int\limits_{V}^{}\!\!d^2 \underline{x} 
\Big[(\part_\perp\gs)^2 + 
(\part_\perp\gp)^2 + 2 \gm^2\gs^2 + \nonumber \\ 
&+2\sqrt{\gl}\gm\gs(\gs^2+\gp^2) + 
\frac{\gl}{2}(\gs^2+\gp^2)^2\Big]
\label{SSBH}
\eea
(see the remark after Eq.(\ref{Tb})). 
The form of the above Hamiltonian suggests that $\gs(x)$ corresponds to a 
massive field because its mass term has a correct sign while the mass term 
is missing for $\gp(x)$ which became a Goldstone boson field. This tree-level  
result is more rigorously expressed by the Goldstone theorem.  

In the usual proof of the Goldstone theorem \cite{GSW}, one inserts a complete 
set of energy and momentum operator eigenstates into the VEV of the commutator 
\beq
[Q,\gp(x)]=\gs (x)
\label{basicc}
\eeq
and then invokes translational invariance to show 
a singularity in the spectral function for $p^2=0$ \cite{GSW,Swieca}. This 
means that there exists a massless state in the spectrum. We can proceed 
analogously because we have all the necessary components for the proof. 
A difference with respect to the usual theory is that here we have an 
explicit realization of the vacuum in the Fock representation, not just an 
abstract state with postulated properties. The states $\vert \ga;v \rangle$ 
represent however only an approximative variational estimate of true 
degenerate family of ground states. But its existence tells us that there 
must exist exact eigenstates of the LF Hamiltonian with energy lower than the 
energy of the Fock vacuum $\vert 0 \rangle$. This is sufficient for the usual 
proof of the Goldstone theorem. Some ingredients of the proof are 
actually valid also if we used the approximative $\vert \ga;v\rangle$ states. 
Namely, the above commutator (\ref{basicc}) is a rigorous consequence of 
Eqs.(\ref{rotsipi}) and (\ref{shift2}). To show that, one only has to use the 
infinitesimal of both transformation laws and compare the leading terms in the 
expansion. The vacuum expectation value of the commutator is   
\beq
\langle v;0 \vert \big[Q,\gp(0) \big] \vert v;0 \rangle = \langle 0 \vert 
U^{-1}_\sigma(v) \gs(0) U_\sigma(v) \vert 0 \rangle = v. 
\label{Gprof}
\eeq
If we denote the set of exact vacuum states by $\vert \Omega_\ga\rangle$, then 
we should also have 
\beq
\langle \Omega_0\vert \big[Q,\gp(0)\big]\vert \Omega_0\rangle = \langle 
\Omega_0 \vert \gs(0)\vert \Omega_0\rangle = f_v,
\label{trueg}
\eeq
where $f_v$ is the expectation value (not known precisely) of the $\gs$ field 
in the true physical vacuum $\vert \Omega_\ga \rangle$. Let $\vert n \rangle$ 
be the set of simultaneous eigenstates of the LF energy and momentum operators, $P^\mu \vert n \rangle = p^\mu \vert n \rangle$, where $p^\mu = (E^-_n,P^+_n,
P^1_n)$.    
Inserting such a complete set in the form of $\hat{1} = \sum_n \vert n 
\rangle \langle n \vert$ into the relation (\ref{trueg}), using the definition 
of the charge operator as a volume integral of $j^+(x)$ as well as the 
translational invariance of the theory, 
\beq 
j^+(x)=\exp{(ix_\mu P^\mu)}j^+(0)\exp{(-ix_\mu P^\mu)},~~
\exp{\big(ix_\mu P^\mu\big)}\vert \Omega_\ga \rangle = \vert \Omega_\ga 
\rangle,
\label{transl}
\eeq
we find
\bea
\frac{2}{V}\sum_{n}\gd^2(\ulip_n)\exp{\big(-\frac{i}{2}E^-_n\xpl\big)}\langle 
\Omega_0 \vert j^+(0)\vert n \rangle\langle n \vert \gp(0)\vert \Omega_0 
\rangle - 
\nonumber \\ 
- \frac{2}{V}\sum_n\gd^2(\ulip_n)\exp{\big(\frac {i}{2}E^-_n\xpl\big)}\langle  
\Omega_0\vert \gp(0)
\vert n \rangle \langle n\vert j^+(0)\vert \Omega_0 \rangle = f_v.
\label{proof}
\eea 
It follows from the VEV of the volume integral of the commutator 
$[\delmu j^\mu,\gp(0)] = 0$ that $f_v$ has indeed to be $\xpl$-independent:
\beq
\Big[\Big(\delp \int\limits_{V}^{}\!\!d^2 \underline{x}j^+(x) + 
\int\limits_{V}^{}\!\!d^2 \underline{x}\delm j^-(x) +  
\int\limits_{V}^{}\!\!d^2 \underline{x}\partial_\perp 
j^\perp(x)\Big),
\gp(0)\Big] = 0, 
\label{tindep}
\eeq
where the second and the third term in the commutator vanishes due to the fact 
that the current obeys periodic BC in $\xmin$ and $x_\perp$. In order that 
the left-hand side of the equation (\ref{proof}) is also $\xpl$-independent,   
there must exist an eigenstate $\vert G\rangle$ of $P^\mu$ which for 
$p^+=0,~ p^\perp=0$ has $E^-=0$ (so that the $\xpl$- dependence vanishes), 
while $\langle \Omega_0 \vert \gp(0)\vert G\rangle \neq 0,~\langle \Omega_0 
\vert j^+(0)\vert G\rangle \neq 0$. Since $M^2=E^-p^+ - p_\perp^2$, this 
state is massless. Note that the Nambu-Goldstone state is not  simply 
$c^\dagger(\underline{k}) \vert \Omega_0 \rangle$ since the latter is not an 
eigenstate of $P^-$. The correct linear combination of Fock states can be 
(at least in principle) obtained by a Hamilton matrix diagonalization.   

\section{Conclusions}

To summarize, in this work a novel strategy to the spontaneous symmetry 
breaking phenomenon in the ligh front description was formulated.  
The approach is based on quantization in a finite volume and on a unitary 
transformation of the Fock LF vacuum to the ground states with lower value of 
the LF energy. These semiclassical vacua are degenerate and have a form of 
boson coherent states. 
The general properties of a spontaneously broken phase of the theory including 
existence of the massless Goldstone boson have been derived in the Fock 
representation. We believe that the present picture of 
spontaneous symmetry breaking in light front field theory in terms of 
semiclassical vacuum states adds a further evidence that there is no conflict  
between the ``triviality'' of the LF vacuum of interacting models and a rich 
nonperturbative contents of quantum field theory.    

\section{Acknowledgement}

This work was supported by the grant No. APVT 51-005704 of the Slovak Research 
and Development Agency and by the VEGA grant No. 2/6068/2006.
The author thanks A. Harindranath and K. Yamawaki for useful discussions. 
Financial support of the Japan Society for the Promotion of Science during 
the author's stay at the Nagoya University and hospitality of the Nuclear 
Theory Group at the Iowa State University is also gratefully acknowledged. 

\newpage

\section{Appendix}

We present a few details of the regularized Dirac delta function and the sign  
function in this Appendix for completeness. Regularization is performed in two 
steps: a cutoff on number of modes (as discussed in the main text) and a 
convergence factor governed by a small parameter $\ge$. The corresponding 
formulae read
\bea
&&\gd_\gL(\xmin - \ymin) = \frac{1}{2L}\sum_{n=1/2}^{\gL}\Big(
e^{-\frac{i}{2}p^+_n(\xmin-\ymin -i\ge)} + 
e^{\frac{i}{2}p^+_n(\xmin-\ymin +i\ge)}\Big), \nonumber \\  
&&\ge_\gL(\xmin - \ymin) = \frac{4i}{L}\sum_{n=1/2}^{\gL}\frac{1}{p^+_n}\Big(
e^{-\frac{i}{2}p^+_n(\xmin-\ymin -i\ge)} - 
e^{\frac{i}{2}p^+_n(\xmin-\ymin +i\ge)}\Big). 
\label{deleps}
\eea
The $\pm i\ge$ terms in the exponents ensure a smooth behaviour in the 
neighbourhood of the points where these functions diverge (for $\gL 
\rightarrow \infty$) or drop to zero. This is quite analogous to the continuum 
theory where the same convergence factors guarantee existence of 
corresponding integrals that replace the discrete series (\ref{deleps}) 
\cite{LMML}. The figures display differences between the functions with 
and without the convergence factors for typical values of the box length 
and of the number of field modes. The shifted function $\ge_\gL(L-\xmin)$ is 
equal to unity to a very high precision over the whole interval $\vert \xmin 
\vert \le L$ except for the endpoints $\xmin =\pm L$ where it behaves in the 
same manner as $\ge_\gL(\xmin)$ in the neighbourhood of $\xmin =0$.
    
\begin{figure}
\centering
\epsfig{figure=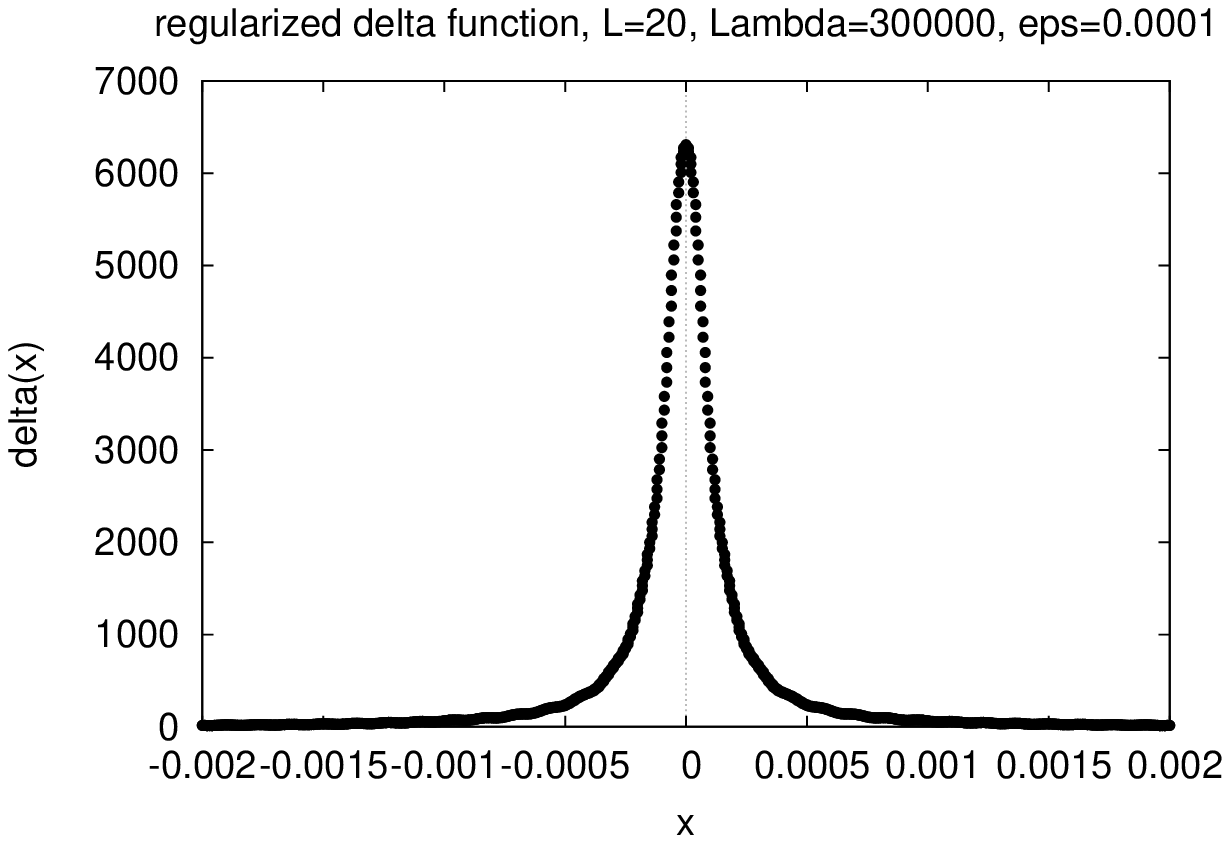,width=100mm,height=70mm}
\caption{Regularized delta function $\gd_\gL(\xmin)$ for $L=20, 
\gL = 3.10^5$ and $\ge = 10^{-4}$.}
\label{rdel}
\end{figure}

\begin{figure}
\centering
\epsfig{figure=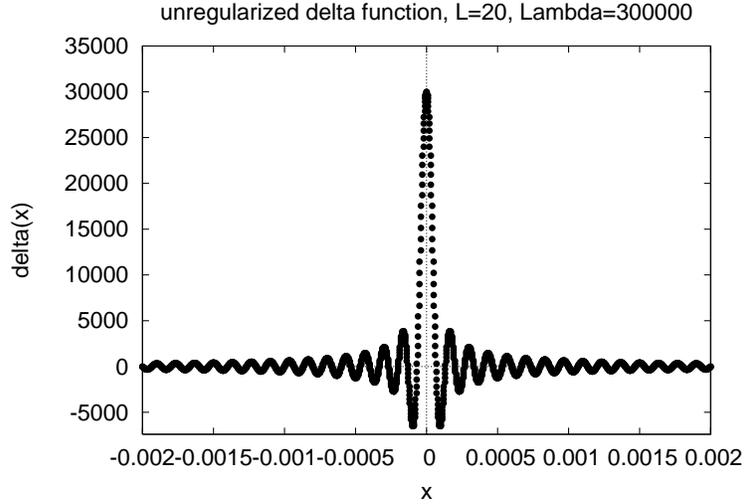,width=100mm,height=70mm}
\caption{Unregularized delta function $\gd_\gL(\xmin)$ (the parameter $\ge=0$) 
for $L=20$ and $\gL = 3.10^5$. The oscillations are rather strong and do not 
vanish for increased values of $\gL$.}
\label{urdel}
\end{figure}

\begin{figure}
\centering
\epsfig{figure=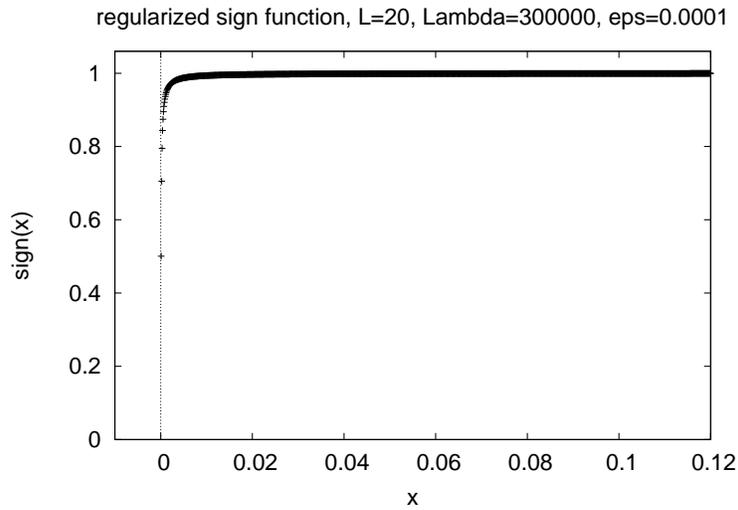,width=100mm,height=70mm}
\caption{Regularized sign function $\ge_{\gL}(\xmin)$ for $L=20, \gL=3.10^5$   
and $\ge = 10^{-4}$ in the neighbourhood of $\xmin=0$. The function 
is a constant to a very high precision over the whole interval except for a 
tiny neighbourhood of the point $\xmin=0$ (or of the edpoints $\xmin = \pm L$ 
in the case of the shifted function $\ge(L-\xmin)$).} 
\label{reps}
\end{figure}

\begin{figure}
\centering
\epsfig{figure=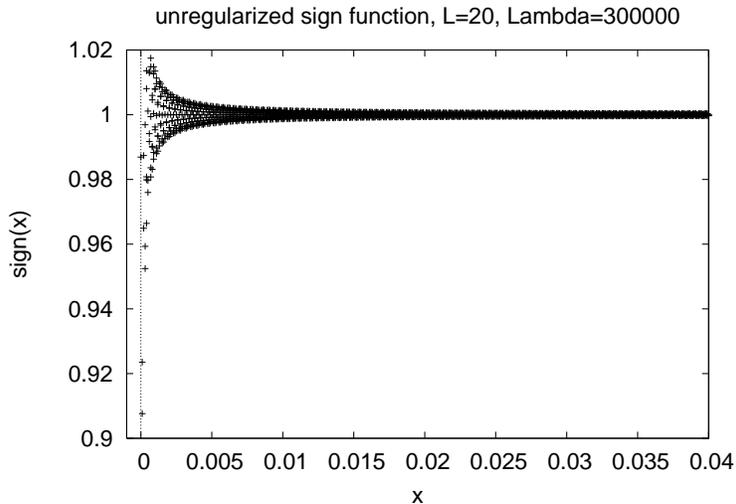,width=100mm,height=70mm}
\caption{Detailed behaviour of the unregularized ($\ge=0$) sign function  
$\ge_{\gL}(\xmin)$ for $L=20, \gL = 3.10^5$ around  
$\xmin=0$. Although the function is again indistinguishable from a constant 
inside the finite interval, it oscillates around the point $\xmin=0$ in 
contrast to the case with a non-zero regulator $\ge$.}
\label{ureps}
\end{figure}
 
\end{document}